# Thermodynamic State of the Interface during Acoustic Cavitation in Lipid Suspensions

*Shamit Shrivastava and Robin O. Cleveland*

*Department of Engineering Science, University of Oxford, UK*

**Abstract:** The thermodynamic state of lipid interfaces was observed during shock wave induced cavitation in water with sub-microsecond resolution, using the emission spectra of hydration-sensitive fluorescent probes co-localized at the interface. The experiments show that the cavitation threshold is lowest near a phase transition of the lipid interface. The cavitation collapse time and the maximum state change during cavitation are found to be a function of both the driving pressure and the initial state of the lipid interface. The experiments show dehydration and crystallization of lipids during the expansion phase of cavitation, suggesting that the heat of vaporization is absorbed from within the interface, which is adiabatically uncoupled from the free water. The study underlines the critical role of the thermodynamic state of the interface in cavitation dynamics, which has mechanistic implications for ultrasound-mediated drug delivery, acoustic nerve stimulation, ultrasound contrast agents, and the nucleation of ice during cavitation.

**Introduction**

Cavitation is the formation of vapor cavities in a liquid due to a tensile stress (1) and is assumed to be one of the key mechanisms for biophysical effects of ultrasound and in particular shock waves (2, 3). It has been suggested as a possible mechanism both in transmembrane and intracellular drug delivery (4) as well as acoustic excitation of the peripheral nervous system (5). The key biophysical system in both these cases is the cellular membrane, which is essentially a self-assembled macromolecular bilayer mainly composed of lipids. To determine the dynamics of the membrane during cavitation it is necessary to monitor the membrane with microsecond temporal resolution. Direct high speed imaging provides morphological images of cell deformation (6) but not the thermodynamic state changes that can be achieved through fluorescence dyes (7). Fluorescent dyes have played a fundamental role in explaining native dynamic processes in biological membranes, such as membrane structural changes during a nerve impulse(8). Here a fluorescent dye is employed to probe the dynamics of aqueous lipid membranes resulting from cavitation.

Lipid surfactants provide a way of controlling the thermodynamic state of an interface and have been investigated in great detail, partly because of their ubiquity in biological systems(9, 10). Thermodynamically lipid membranes can be treated as quasi-2D systems, where the state can be defined by surface observables (e.g. area/molecule) and corresponding thermodynamic fields (e.g. surface pressure) (11). Molecules which co-localize at the interface, such as fluorescent probes, can be added to the system and used to indicate the thermodynamic state of the interface. For example, we recently showed that the wavelength shift of the emission spectrum of the fluorescent probe Laurdan can be used to dynamically measure the thermodynamic state of lipid interfaces (7), when subjected to microseconds width acoustic impulses.

The enthalpy of the vapor-liquid interface during cavitation can be expressed as $\Delta H_i = T_i \Delta S_i + A \Delta \sigma$, where $\sigma$ is the surface tension, $A$ is surface area, $T_i$ is the interfacial temperature, and $S_i$ is the interfacial entropy (12, 13). At the formation of a cavity in water under normal pressure and temperature, $\Delta \sigma$ is typically of the order of 72mN/m, giving a theoretical tensile strength or the cavitation threshold of water in excess of 100MPa (14). However, the tensile strength of tap water is practically of the order of 0.1 MPa (15) which can be attributed to the presence of particles with crack and crevices that stabilize cavitation nuclei (16) or to impurities acting as surfactants that lower the surface tension. The latter case is relevant to this work, as the cavitation threshold has been shown to be significantly lower for fat-water or oil-water interfaces than in either pure water or pure oil (17), and cavitation has been observed to initiate specifically at the interfaces (1). Subsequent macroscopic expansion of the cavity in the presence of surfactants has been studied widely in the context of lipid coated bubbles, see for example the review by Doinikov et.al. (18). Thus the surfactants can affect not only the nucleation but also the subsequent macroscopic evolution, which several models have tried to capture by allowing for a radius dependent surface tension, $\sigma(R)$ (19) in the Rayleigh-Plesset equations, and making the assumption that the energy of the interface is given by $\Delta E_i = \sigma \Delta A$. However, this ignores the entropy or the heat content of the interface. The complete expression is $\Delta E_i = \sigma \Delta A + T_i \Delta S_i$ (13), where $\Delta S_i = -\frac{d\sigma(T)}{dT} \Delta A$ (13) and this study will show that the entropy term is significant and is required to explain the observed role of the interface during cavitation even qualitatively.

Here we investigate cavitation in a suspension of lipid vesicles by observing the spectral shift in the fluorescence emission of a hydrophobic probe Laurdan, co-localized at the interface (7) with sub-microsecond resolution. The lipid composition and the temperature of the suspension were varied to change the initial state of the lipid interfaces. The cavitation was generated by firing focused acoustic shock waves into the aqueous suspension. No a priori assumptions about the site or the spatial details of the evolving cavitation field were made, rather they were *deduced* from the experimental observations presented in this study.

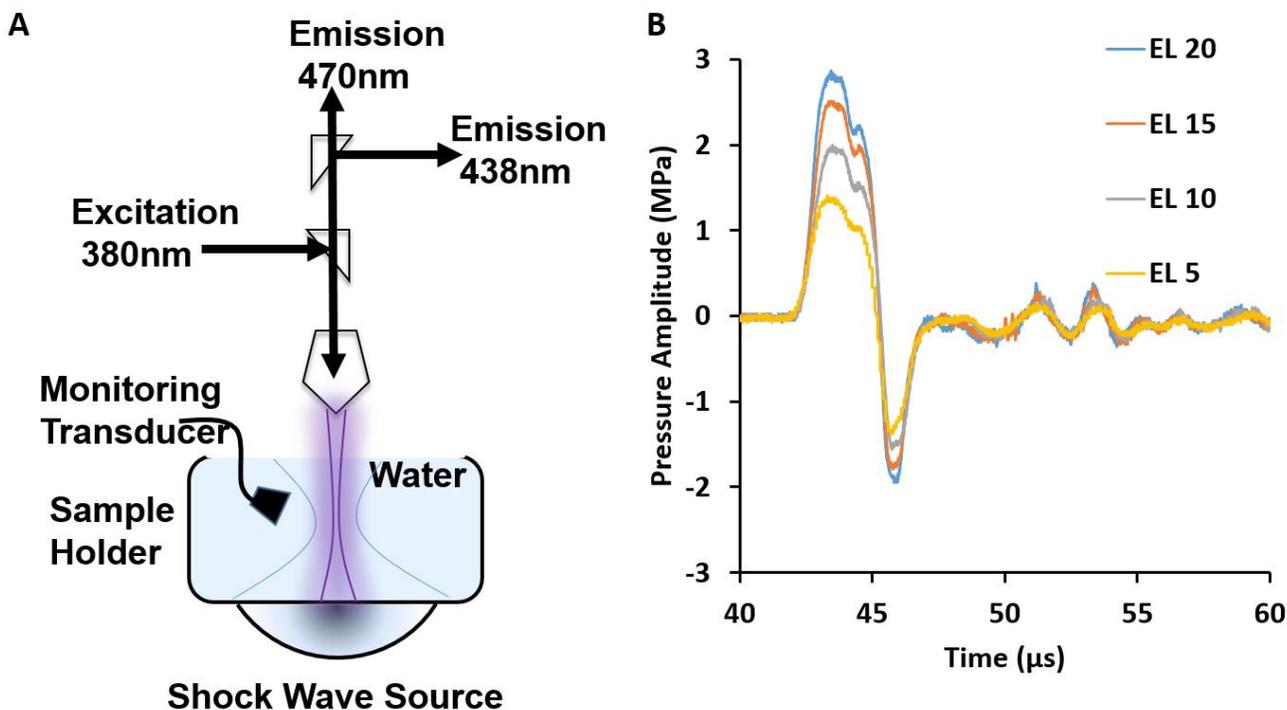

**Figure 1 Experimental Setup (A)** The shock wave source is coupled to a temperature controlled water tank. The optical system consists of a LED light source (380 nm) which is focused by an objective in the water, coincident with the shock source, and fluorescent signals are detected by two photomultiplier tubes recording simultaneously at two different wavelengths (438 and 470nm). An unfocused monitoring transducer is used to detect acoustic emissions from the vicinity of the focal region. During experiments multilamellar lipid vesicles (MLVs), embedded with the dye Laurdan, are present everywhere in the tank. (B) Measured focal pressure waveforms for shock waves at four different energy settings.

**Material and Methods**

The experimental setup and the MLV preparation, have been described in detail previously (7). The lipids 1,2-dioleoyl-sn-glycero-3-phosphocholine (DOPC 850375), 1,2-dimyristoyl-sn-glycero-3-phosphocholine (DMPC 850345), 1,2-dipalmitoyl-sn-glycero-3-phosphocholine (DPPC 850355) were purchased as a 25 mg/mL solution in chloroform from Avanti Polar Lipids, Inc. (Alabaster, AL, USA). The flourescent probe Laurdan was purchased from Thermo Fisher Scientific, (Waltham, MA, USA). MLVs were prepared from 2.5mg of one of the lipids (DOPC, DMPC or DPPC) and $50\mu g$ of Laurdan and added to a temperature controlled water tank containing 500ml of deionized water, equilibrated to the experimental temperature. For determining the cavitation threshold 12.5 mg of lipids were extruded through a 100nm polycarbonate filter at 60°C and added to 500mL of deionized (18 MOhms resistance) and degassed water. The lipid nanoparticles were sized using differential light scattering (Malvern Panalytical, Zeta Sizer, Spectris USA). At 24°C and pH 7, the mean hydrodynamic radius and poly dispersity index were as follows: DOPC (160nm, 0.2), DMPC (140nm, 0.17) and DPPC (452nm, 0.44).

The phase state of each lipid was characterized in terms of a "reduced temperature", $T^* = \frac{T-T_m}{T_m}$, where $T_m$ is the temperature of the main phase transition, equivalent to the melting point of of the lipid system. Therefore when $T^*<0$ the lipid is in the $L_\beta$ gel phase, when $T^*>0$ it is in the $L_\alpha$ fluid phase, and when $T^* = 0$ it is at transition.

Shock waves generated by a Swiss Piezoclast® (EMS Electro Medical Systems S.A, Switzerland) were fired in to a small test chamber that had been filled with the lipid solution, and focused 1 cm below the free surface (Fig 1A). The pressure waveforms generated by the Piezoclast were measured with a PVDF needle hydrophone (Muller-Platte needle probe, Dr. Muller instruments, Oberursel, Germany) with a manufacturer specified sensitivity of 12.5mV/MPa. The needle was placed at the focus of the Piezoclast, i.e. 1 cm below the free surface, and measured shockwaves at four different energy levels are shown in fig.1B. The waveforms exhibit a leading positive pressure followed by negative tail and the peak pressure was of the order of 3MPa peak-to-peak. Acoustic emissions were monitored by using an immersed unfocused ultrasound transducer with 10 MHz central frequency and 3.175 mm diameter (V129-RM, Olympus, Waltham, MA). The monitoring transducer was placed at an angle of 45° from the axis of the shock wave, see Fig 1A, at a distance of $\sim 1.5 cm$ from the focal region. The emissions were digitally filtered using a 7-20MHz band pass filter (3$^{rd}$ order Butterworth), to attenuate the signal from the shock source and its reflections. Labview 2015, (National Instruments, Austin, TX, USA) was used for all the data acquisition and analysis requirements.

A custom built upright epifluorescent microscope, with an objective with 1.3 cm working distance (LMPLFLN 20X Objective, Olympus, Tokyo, Japan), was placed above the free surface of the test chamber and the focus of the objective was aligned with the focus of the shock waves (Fig 1A). The optical field of view (FOV) was of the order of $400 \mu m \times 400 \mu m$. For fluorescence detection, the FOV was illuminated by a UV LED (central wavelength 385nm and bandwidth 10nm, catalog# M385LP1, Thorlabs, Newton, NJ USA) and the emitted fluorescence intensity was measured simultaneously by two photomultiplier tubes (HT493-003, Hamamatsu, Japan), $I_{438nm}$ (filter at $438 \pm 12\ nm$) and $I_{470nm}$ (filter at $470 \pm 11\ nm$), with a signal bandwidth of 8MHz.

The measured optical signal was evaluated as $\frac{\Delta I}{I_o} = \frac{I(t)-I_o}{I_o}$, where $I_o$ is the initial emission intensity and it depends on the thermodynamic state of the interface, the concentration of dye molecules and the design characteristics of the optical setup, e.g. focal volume. The rate of mechanical deformation induced by the shock wave was characterized by $\dot\gamma = \frac{d}{dt}\frac{\Delta I_{470nm}}{I_{470nm}}$. The thermodynamic state of the membrane was characterized by $\frac{\Delta RP}{RP_0} = \frac{\Delta I_{438nm}}{I_{438nm}} - \frac{\Delta I_{470nm}}{I_{470nm}}$ which due to normalization is insensitive to the amplitude of the light intensity (7). For small perturbations $\frac{\Delta RP}{RP_0} \propto \Delta h_i$, i.e. change in the enthalpy of the interface (7).

**Results**

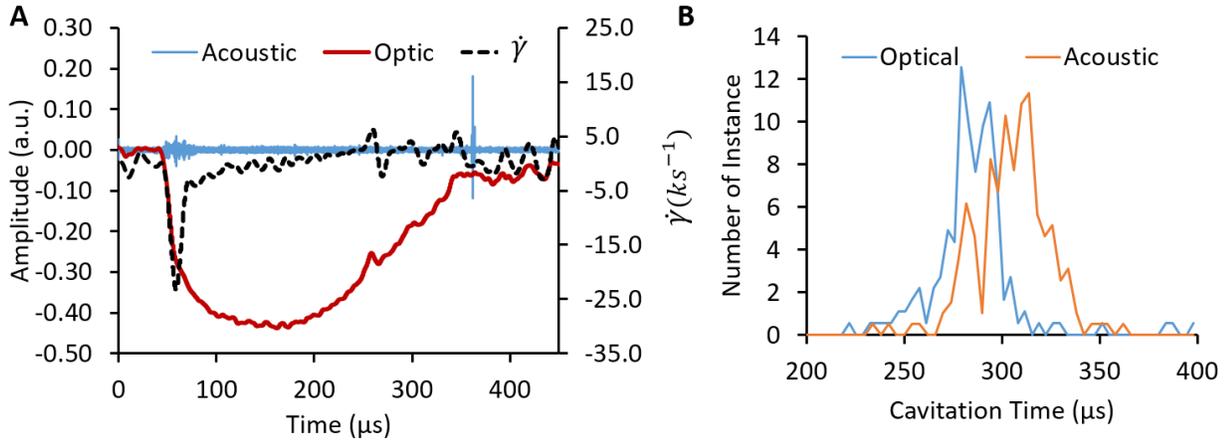

**Figure 2** (A) Optical intensity $\frac{\Delta I}{I}$ at 470 nm, acoustic emissions, and rate of deformation $\dot{\gamma}$ measured during a single cavitation event in DOPC suspension at 20C in response to a shock impulse of energy level 20. The shock wave arrives at 50 µs resulting in an acoustic emission, a drop in optical intensity and a peak of rate of deformation. The cavitation bubble/cluster grows to a maximum extent at 170 µs and then collapses at around 350 µs producing a strong acoustic emission and a knee in the optical intensity. (B) The distribution of cavitation time for 200 such shock waves as determined from the acoustic and optical signals.

Figure 2 shows the acoustic emission, optical signal ($\frac{\Delta I}{I_0}$), and the mechanical deformation rate ($\dot{\gamma}$) measured from a DOPC suspension in response to a single shock wave. The shock wave was triggered at $t = 0\mu s$ and arrived at the focus at around $t_0 = 50\mu s$, resulting in: (i) a rapid drop in the optical signal, indicating formation of bubbles, (ii) an acoustic emission, which was detected at 60 µs due to the 10 µs propagation time to the monitoring transducer, and (iii) a negative spike in the rate of deformation. The optical signal continued to drop until about $170\mu s$ after which it started to increase until it "recovered" at $t_r = 350\mu s$, although not to baseline. The optical signal $\frac{\Delta I}{I_0}$ is consistent with Mie scattering based optical measurements during similar cavitation experiments (see e.g. figure 7 in ref. (20)), including the incomplete recovery of optical signal at $t_r = 350\mu s$, indicating that $\frac{\Delta I}{I_0}$ is related to the size of the bubble cluster. The time of recovery corresponds to a second stronger acoustic emission at $t = 360\mu s$, which is consistent with the collapse of a bubble or bubbles (20). This double acoustic signature is characteristic of the creation and collapse of a bubble or bubble cluster in response to a shock wave and the time between the signals, $t_c$ is an indicator of the intensity of the cavitation (21, 22). The lack of a distinct peak in $\dot{\gamma}$ at recovery, similar to the one observed at $t_0 = 50\mu s$, suggests that the observed decay in fluorescence signal is most likely dominated by lipid relaxation and diffusion, which is different from the collapsing cavity.

The cavitation time $t_c = t_r - t_0$ was determined from the optical signals where $t_0 = 50\mu s$ is the arrival time of the shock wave and $t_r$ was defined as the time when $\frac{\Delta I}{I_0}$ returned above a value of $-0.1$, in the time window t>90 μs to ensure the initial emission was complete. If $\frac{\Delta I}{I_0} < -0.1$ at $t = 90\mu s$ then it was assumed that there was no detectable cavitation. As shown in fig. 2B, both optical and acoustic detection provides a consistent estimate of cavitation time and in what follows only optically observed cavitation times have been reported.

In fig. 3, the cavitation time is reported as a function of shock wave energy for MLVs in four different initial states, as indicated by the corresponding reduced temperature $T^*$. In all cases, cavitation time increases monotonically with increasing energy level. This is consistent with shock wave induced cavitation measured in water (21). For the two formulations in a condensed state $(T^* < 0)$, $t_c < 200\mu s$ on average, whereas for the two formulations in a fluid state $(T^* > 0)$, $t_c > 250\mu s$. This suggests that the duration of the cavitation is dependent on the thermodynamic state of the lipid membrane, with the longer cavitation times observed for more fluid initial states. Note that from the number of cavitation events plotted in fig.3, the presence of significant cavitation at EL 5 and 8 for lipids close to gel – fluid phase transitions ($T^* = -0.045$ and $T^* = 0.046$), indicating the cavitation threshold is lower near the transition (SI Fig. S1). Therefore, this effect was investigated in more detail.

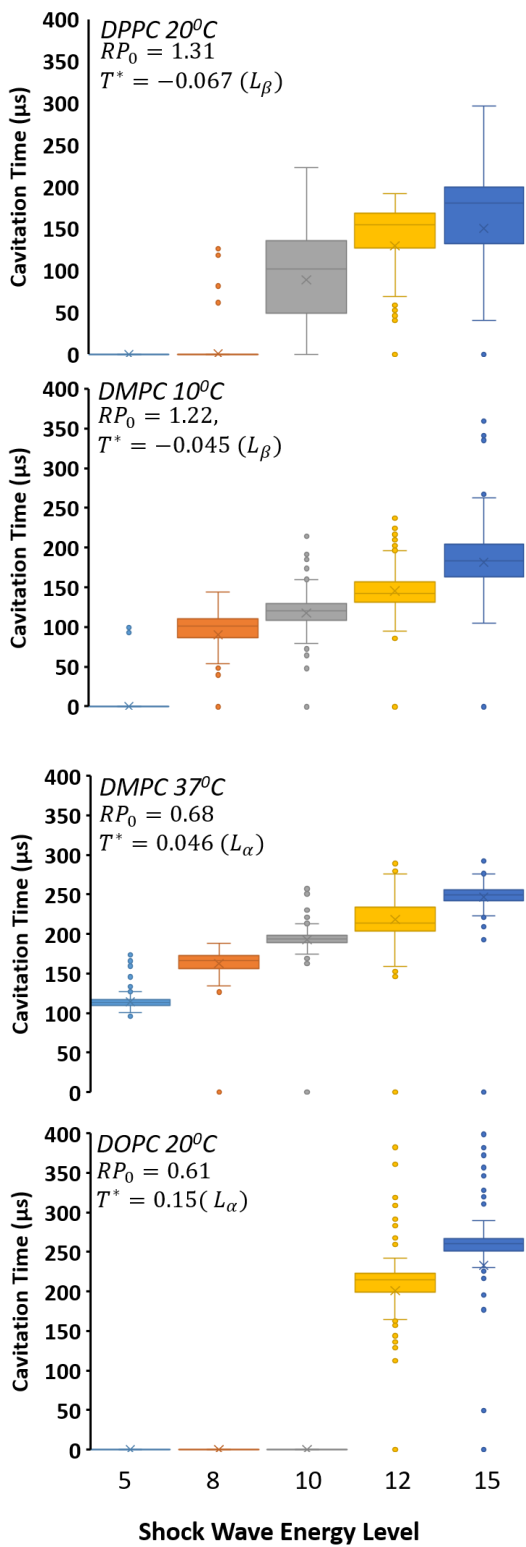

**Figure 3** Dependence of cavitation time as a function of shock wave energy level for four different initial lipid states.   The initial state ranges from the condensed state, T*=-0.067 (top) to fluid state, T*=0.15 (bottom). A cavitation time of zero indicates the absence of cavitation. For the gel initial state T*=0.15, cavitation starts at EL 10 and the cavitation time increases with EL to a maximum of 200 µs. At T*=-0.045 significant cavitation occurs for EL 8 and cavitation time increases with EL but again doesn't exceed 200 µs. At T*=0.046, the threshold is lowered and cavitation occurs at all energy levels settings with a maximum cavitation time of 250 μs. For T*=0.15 cavitation starts at EL 12 and the cavitation time exceeds 250 μs at the highest setting. Dots indicate the outliers that are more than 1.5 times the upper/lower quartile.

The effect of the thermodynamic state of lipid nanoparticles on cavitation threshold was determined by determining the cavitation rate (fraction of shock waves that result in detectable cavitation) as a function of energy level. Figure 4 shows the effect of different lipids at the same temperature and pressure on the cavitation threshold. The cavitation rate was measured based on 100 shocks at 5 different energy levels and dose response fits were obtained. For each preparation the energy level was ramped from 5 to 15 at two different PRFs of 1 Hz and 3 Hz. It can be seen that at both PRFs, the lipid preparation with $T^*$ closest to zero, i.e. DMPC, has the lowest threshold; further supporting the concept that the nucleation threshold is indeed lowest when the lipid is close to a phase transition.

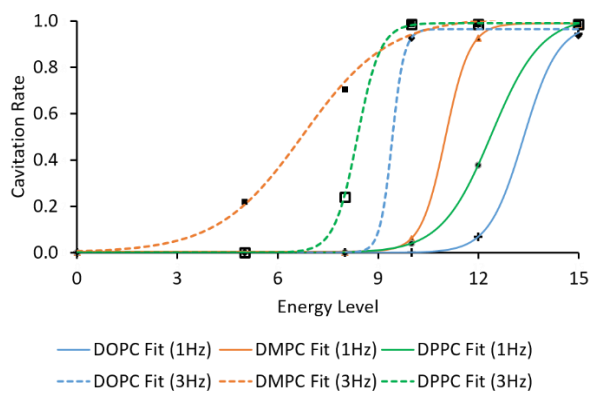

**Figure 4** Cavitation rate of aqueous suspension of liposomes measured as a function of the energy level, the thermodynamic state and the repetition rate. Cavitation rate defined as number of cavitation events per 100 shocks have been plotted at two different shock repetition rates 1Hz (solid curves) and 3Hz (dashed curves). Three different liposome suspensions were tested. The mean hydrodynamic radius and poly dispersity index were as follows: DOPC (160nm, 0.2), DMPC (140nm, 0.17) and DPPC (452nm, 0.44). All samples were tested at a temperature of 24 °C and pH 7.

Figure 5 shows the evolution of $\frac{\Delta RP}{RP_0}$ and $\frac{\Delta I}{I_0}$ for the four different initial states. Due to the low signal-to-noise ratio the traces are averaged over 500 shocks. Note the shock waves energy levels are not the same but correspond to the energy level at which "consistent cavitation" was observed; defined here as obtaining single peak in $t_c$ histogram and $|\dot{\gamma}| > 3 \times 10^{-4} s^{-1}$. It can be seen that for all initial states, $\frac{\Delta RP}{RP_0}$ and $\frac{\Delta I}{I_0}$ have similar dynamics, that is, they are responding to the growth and collapse of cavitation.

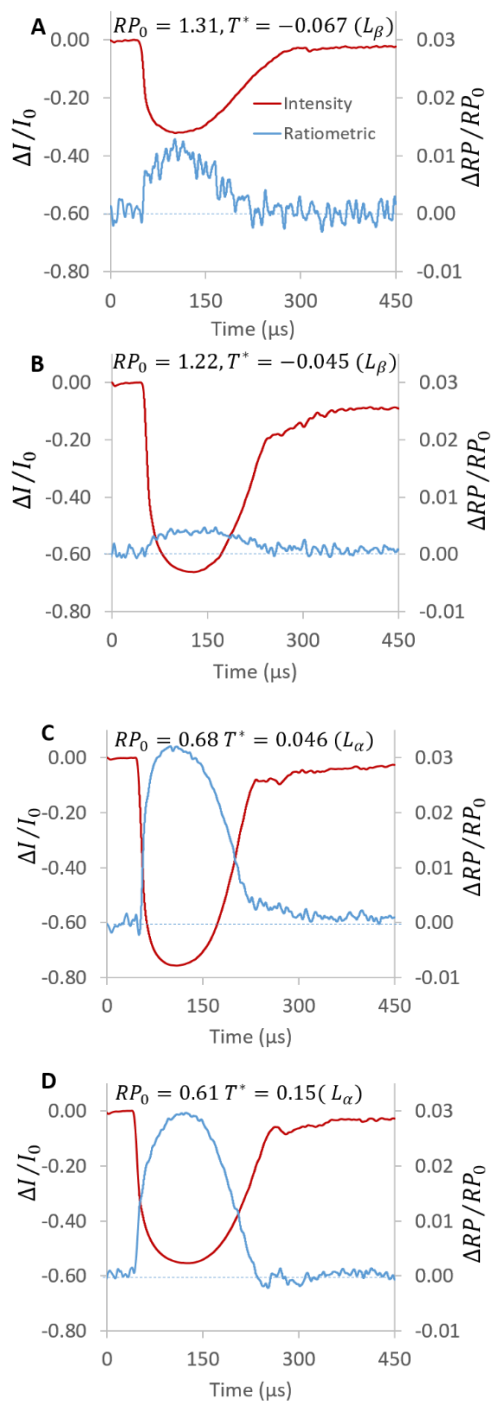

**Figure 5** Optical intensity $\frac{\Delta I}{I}$ and ratiometric parameter $\frac{\Delta RP}{RP_0}$ are plotted as a function of time for four initial conditions. Lipids and temperature give the initial condition, which is also reported by $RP_0$, These are A) DPPC T=20℃ EL 15 B) DMPC, T=10℃, EL 15 C) DMPC, T=37℃, EL 10 D) DOPC, T=20℃, EL 12. The optical intensity change is greatest when $T^*$ is closest to zero indicating that the cavitation bubbles/cluster attains the greatest volume when the lipid is close to a transition. $\frac{\Delta RP}{RP_0}$ is greater for lipids in fluid state than gel state indicating a greater state change.

However, for an initial gel state ($T^* < 0$) the peak is lower with $\frac{\Delta RP}{RP_0} < 0.15$ whereas for initial fluid state ($T^* > 0$), the peak $\frac{\Delta RP}{RP_0}$ is much larger ($\frac{\Delta RP}{RP_0} \approx 0.3$). Based on the interpretation of Laurdan spectrum, which will be discussed below, a $\frac{\Delta RP}{RP_0} > 0$ corresponds to condensation of the lipid interface. Therefore, these observations suggest that during the expansion of the cavitation cluster, the lipids undergo condensation. Furthermore, the condensation is more pronounced the lipids are initially in a fluid state. The significance of the quantitative differences in the response as a function of $T^*$ will be discussed in more details below. Note that $\left|\frac{\Delta I}{I_0}\right|$ does not capture these state differences, as the signal is dominated by the appearance of gaseous scatterers in the FOV. In fact *to a first order*, the trend with $\left|\frac{\Delta I}{I_0}\right|$ would indicate that the fraction of such scatter is minimum for $T^* = -0.067$, $\left|\frac{\Delta I}{I_0}\right| \approx 0.3$; peaks for $T^* = 0.045$, $\left|\frac{\Delta I}{I_0}\right| \approx 0.75$; and then decreases again for $T^* = 0.15$, $\left|\frac{\Delta I}{I_0}\right| \approx 0.55$. This is consistent with the observed lowering of cavitation threshold near a phase transition, which would imply more nucleation sites for the given field of view.

As the shock wave energy was increased there was an increase in $\dot{\gamma}$ as shown in Fig. 6A and B. This is consistent with a more violent expansion of the cavity. However, the peak $\left|\frac{\Delta RP}{RP_0}\right|$ remained at ~0.15, suggesting a saturation of the change in the thermodynamic state of the interface during rapid expansion; a phenomenon which is common in such nonlinear systems (23).

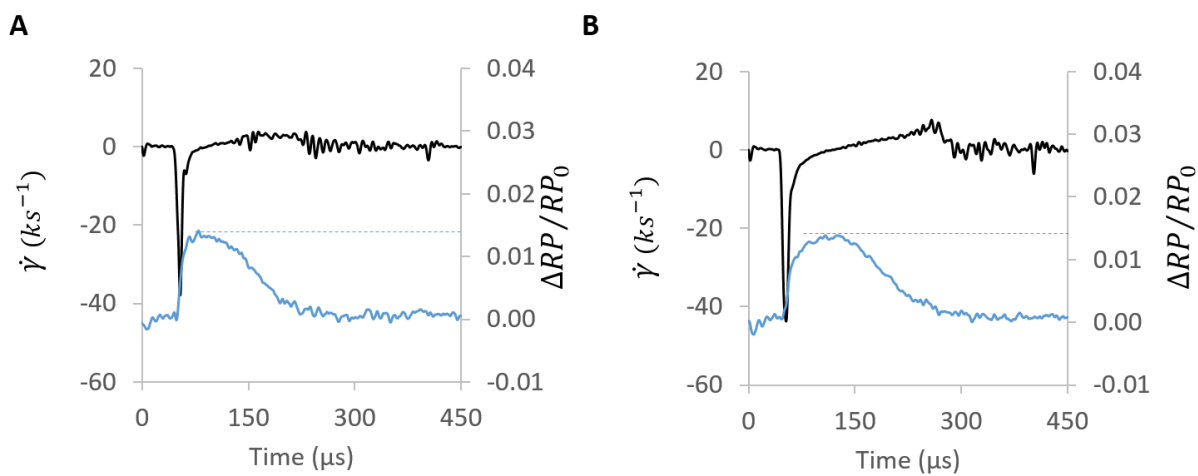

**Figure6** $\Delta RP/RP_0$ and $\dot{\gamma}$ averaged over 500 shocks at energy levels (A) 12 and (B) 15 for DMPC suspension at 23°C. At the higher energy level the growth rate $\dot{\gamma}$ was larger indicating the cavitation was driven harder; but the $\Delta RP/RP_0$ plateaued to the same value suggesting the lipid obtained the same condensed state.

A bimodal distribution in the cavitation time $t_c$ was often observed for EL 12 and 15, an example is shown in SI Fig 2, suggests that the cavitation occurred in two distinct paths in the state diagram. To examine the difference in the two paths, the set of EL 12 waveforms shown in SI Fig 2 were divided into two populations by using a cutoff of $t_r = 270 \mu s$. Figure 7A and 7B shown averaged traces for $\frac{\Delta RP}{RP_0}$ and $\dot{\gamma}$ for the two populations. It can be seen for $t_r > 270 \mu s$ that $\frac{\Delta RP}{RP_0}$ is similar to Fig 5 and the increase in $\frac{\Delta RP}{RP_0}$ suggest the lipid is being condensed. However, quite remarkably for $t_r < 270 \mu s$ $\frac{\Delta RP}{RP_0}$ has the opposite sign suggesting that the lipid is becoming more fluid; it can be seen that this correlates with a relatively low $\dot{\gamma}$. These results suggest that when the cavity has a high expansion rate it condenses the lipid interface $\left(\frac{\Delta RP}{RP_0} > 0\right)$, but when it has a low expansion rate the lipid interface is fluidized $\left(\frac{\Delta RP}{RP_0} < 0\right)$. This insight is only available through the use of Laurdan and quantifying the changes in its emission spectrum it in terms of $\frac{\Delta RP}{RP_0}$.

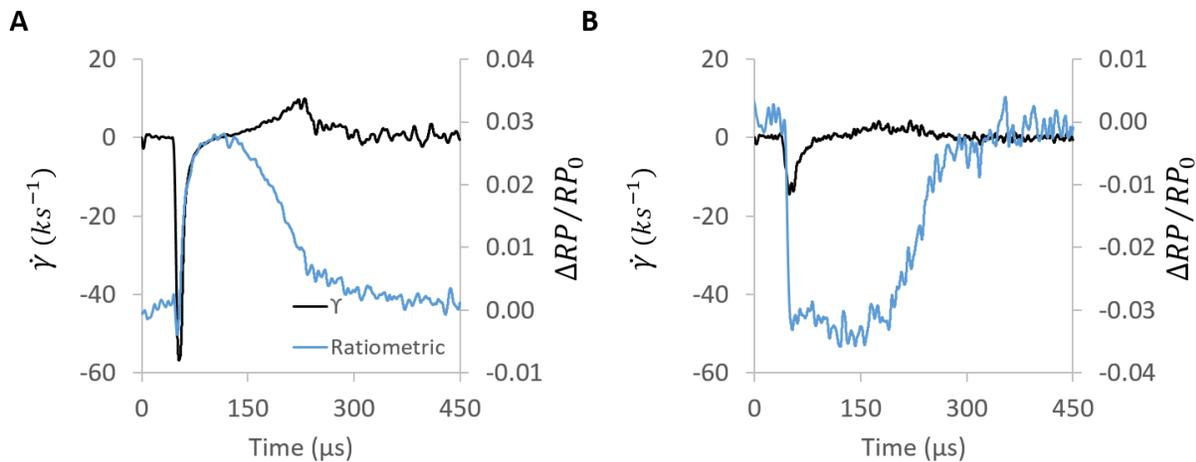

**Figure 7** The rate of mechanical deformation ($\dot{\gamma}$) and $\Delta RP/RP_0$ for a DMPC solution at 37°C (A) waveforms averaged with $t_r > 270 \mu s$ gives high peak $\dot{\gamma}$, and $\frac{\Delta RP}{RP_0} > 0$ indicates the lipid condenses during cavitation (B) waveforms averaged with $t_r < 270 \mu s$ give low peak $\dot{\gamma}$ while $\frac{\Delta RP}{RP_0} < 0$ indicates the lipid fluidizing during cavitation. The distribution of $t_r$ is given in SI Fig 2.

**Discussion**

The experimental results show that for an aqueous-lipid suspension, the characteristic cavitation features such as threshold and cavitation life time, as well as the thermodynamic state of the interface during cavitation, are dependent on the initial state of the lipid interface with the reduced temperature $T^*$ being a strong indicator of nucleation threshold and subsequent evolution of the cavity. Here we interpret these results in terms of thermodynamic principles to arrive at a phenomenological description of the observation and possible microscopic states of the lipid-water interface during cavitation. Our data showed that the cavitation threshold was lowest at phase transition ($T^* \approx 0$) this is not consistent with the usual explanation of cavitation threshold in the presence of lipids or fats, which is attributed to the lowering of surface tension alone. At a given temperature different lipids should have a similar surface tension (24–26). Furthermore, the surface tensions should decreases monotonically with temperature (27), and so a monotonic decrease in threshold would have been expected. Instead we observe a minimum at the phase transition temperature. We propose that the reduced cavitation threshold is because of the maximum in thermodynamic fluctuations that occurs at the phase transitions. Under the assumption of local equilibrium, the amplitude of microscopic fluctuations is related to the second derivative of the thermodynamic potential of the system (See Appendix A). The enthalpy fluctuations of the lipids $|(\delta H_2)^2|$ are given by

$$|(\delta H_2)^2| = k C_{p_2} T^2 \qquad (1)$$

where $k$ is the Boltzmann constant, $T$ is the temperature, and $C_{p_2}$ is the heat capacity of the lipids at constant pressure. The heat capacity peaks at phase transition and so the fluctuations in enthalpy also has a maximum. The fluctuations in the volume of water $V_1$ (liquid *and* vapor) are related to the enthalpy of the lipids $H_2$ (see Appendix A) as

$$|\delta V_1 \delta H_2| \approx \frac{kT^2}{P/C_{p_2} - T \frac{\partial P}{\partial H_2}} \qquad (2)$$

where $P$ is the equilibrium pressure. As described in the Appendix A, the right hand side of eq.(2) also has a maximum at the transition. Therefore a peak in $|(\delta H_2)^2|$ at phase transition due to a peak in $C_{p_2}$ also implies a peak in $|(\delta V_1)^2|$ of interfacial water at the phase transition of the lipid.

Our data is consistent with the hypothesis that the peak in these fluctuations gives rise to the minimum in cavitation threshold. In water, vaporization occurs when pressure drops below saturation limit or the water-vapor boundary in the phase diagram. The vaporization process involves a finite relaxation time and if the pressure drop occurs faster than this relaxation time, such as during a shock wave, the system goes into the metastable regime of the water-vapor phase diagram. In the metastable regime while the system is in mechanical equilibrium with respect to the tensile stress, it is not in chemical equilibrium with respect to the water to vapor transition (partial equilibrium).

Eq. 2 can be extended to the metastable regime under the assumption of local equilibrium (see appendix) where the heat capacity and compressibility of water are defined as functions of the dynamic local state. Therefore, as the system moves further into the metastable regime (increasing heat capacity, compressibility of water), thermal fluctuations also increase eventually causing nucleation, which also results in a local minimum in the P-V curve (spinodal boundary). Beyond this minimum, $\frac{\partial P}{\partial V_1} > 0$ (compressibility of water is negative), which results in thermodynamic instability and sudden *macroscopic* expansion and phase change that takes place even in the absence of an external stress. Eq. 2 is not valid beyond this point as the assumption of reversibility is not valid. The water – to – vapor phase change is accompanied by absorption of latent heat and dissipation due to non-equilibrium heat transfer, which establishes the chemical equilibrium(28). The experimentally observed expansion phase (Fig.2) (starting at t=50$\mu s$) is assumed to represent this process. Therefore, during the observed cavitation expansion the emission spectrum of the dye only measures $\delta H_2$, but $\delta V_1$ cannot be estimated using Eq.2 due to the irreversible nature of the process.

| Lipid | $T_m$ | $T$ | $T^*$ | Initial Phase | $RP_0$ | $\left|\frac{\Delta RP}{RP_0}\right|$ | Phase Change | $T_t$ | $-\Delta h\left(\frac{kJ}{mol}\right)$ |
|---|---|---|---|---|---|---|---|---|---|
| DOPC* | −40°C | 20°C | ~0.15 | $L_\alpha$ | 0.622 | 0.029 | $L_\alpha \to L_c$ | −12°C | ~32 − 65 |
| DMPC | 23.6°C | 37°C | 0.046 | $L_\alpha$ | 0.726 | 0.032 | $L_\alpha \to P_\beta$ | 23.6°C | 25.0 |
|  |  | 20°C | −0.01 | $P_\beta$ | 0.950 | 0.014 | $P_\beta \to L_c$ | 11.4°C | 15.8 |
|  |  | 10°C | −0.045 | $L_\beta$ | 1.208 | 0.004 | $L_\beta \to L_c$ | NA | NA |
| DPPC | 41.5°C | 20°C | −0.067 | $L_\beta$ | 1.300 | 0.012 | $L_\beta \to L_c$ | 18.3°C | 14.2 |

Table 1. Thermodynamic data associated with state changes in different lipid systems as reported in literature (29) and corresponding spectral shift in Laurdan $\left|\frac{\Delta RP}{RP_0}\right|$ as observed in this study. $T_m$ is the chain melting temperature, T is the temperature of the experiment and $T^*$ is the corresponding reduced temperature, all at atmospheric pressure. Possible phase changes from the initial condition and the corresponding transition temperature ($T_t$) as well as enthalpies($-\Delta h$) are listed. (*) A range of values have been reported for phase transitions of DOPC suspension in water due to experimental difficulties in measure subzero transitions in water (30). $L_\beta \to L_c$ in DMPC, on the other hand has not be studied as reliably as the other values reported here, because of the highly metastable nature of the transition (29, 31).

In the data shown here, a rapid expansion of the vapor cavities resulted in condensation of lipids. Previously it has been shown that the coupling between the bulk and interface decreases with decreasing timescales of interactions (32, 33). In our case this implies that if the surrounding water cannot exchange heat with the interface then the heat of vaporization has to come from within the interface. This will result in a decrease in enthalpy of the lipids at the interface and hence their condensation (34, 35). This interpretation is further supported by the observation that at slower expansion rates, when there is sufficient time for the surrounding water to provide heat to the interface, the lipids get fluidized upon increase in the volume (surface area) of the cavity. The state of the lipids will also place constraints on the expansion of the cavity. For example, the cavitation time (Fig. 3) was shortest for the gel state, which presumably limits the amount of heat that can be transferred from the lipids to water for vaporization.

Evidence for the phase transition in lipids is being captured by $\Delta RP/RP_{0(1\to 2)}$ which the experiments show is proportional to corresponding enthalpies of transitions (7, 31, 36), i.e.

$$\Delta RP/RP_{0(1\to 2)} \propto \Delta h_{(1\to 2)} \qquad (3)$$

Table 2 shows the ratio of the peak of the $\Delta RP/RP_0$ and the ratio of the enthalpy of transition (from Table 1). It can be seen that the ratios agree to within about 15% which is consistent with the hypothesis that lipids are indeed condensing to the $L_c$ phase during the expansion phase of cavitation; allowing for the fact that the literature values for enthalpy change are for quasi-static processes. This confirms that $\Delta RP/RP_0$ measures the state changes even during cavitation and not just during small acoustic perturbation as shown before (7). We were unable to identify data in the literature for the $(DMPC)_{(L_\beta \to L_c)}$ enthalpy change, but based on the measurements here we can predict it to be of the order of $-5\frac{kJ}{mol}$.

|  | $(DMPC)_{(P_\beta \to L_c)}$ | $(DPPC)_{(L_\beta \to L_c)}$ | $(DOPC)_{(L_\alpha \to L_c)}$ | $(DMPC)_{(L_\beta \to L_c)}$ |
|---|---|---|---|---|
| $\dfrac{\Delta RP/RP_{0(x)}}{\Delta RP/RP_{0(DMPC)_{(L_\alpha \to L_c)}}}$ | 0.43 | 0.37 | 0.91 | 0.12 |
| $\dfrac{\Delta h_{(x)}}{\Delta h_{(DMPC)_{(L_\alpha \to L_c)}}}$ | 0.38 | 0.36 | 0.71 − 1.43 | NA |

*Table 2. The ratio of maximum amplitudes of the observed $\Delta RP/RP_0$ for different samples were calculated from the values in table 1. Assuming final state as $L_c$, as discussed in the text, corresponding ratio of enthalpies calculated using table 1 are also listed.*

Now to understand the microscopic state of the lipid interface during the process it is important to recall that phase transition in lipid interfaces involves several degrees of freedom with distinct timescales and enthalpy of conformational change, see (36) for a detailed overview. The timescales of the observed spectral shift in this study are of the order of $10 - 100\mu s$ which would correspond to formation of rotational isomers or gauches in the hydrocarbon tails of the lipid molecules and constitute the observed enthalpy change. In the literature a positive $\Delta RP/RP_0$, i.e. a blue shift in the Laurdan spectrum, is usually attributed to an increase in the packing of lipids. Two observations from our data are not consistent with this interpretation. First, we generally saw a blue shift during rapid bubble expansion when one would expect the packing to decrease as the surface area of the bubble increases. Second we observe a saturation effect (recall Fig 6) where cavities driven with stronger shock waves exhibited no increase in $\Delta RP/RP_0$ despite the fact the cavities appeared to grow to a larger size. Instead, the saturation is consistent with the spectral shifts being associated with the thermodynamic state of the lipid. The $\Delta RP/RP_0$ is dominated by the enthalpy associated with the phase transition (table 2), therefore, once lipids have condensed to a gel phase, any further expansion of the bubble would not lead to substantial change in $\Delta RP/RP_0$.

To the best of our knowledge, this is the first time that the role of the thermodynamic state of the interface during cavitation has been investigated. There are two aspects to the role of the thermodynamic state of the interface during acoustic cavitation; (i) the microscopic or statistical physics of nucleation of a vapor cavity at an interface and its effect on cavitation threshold, and (ii) the macroscopic state of the interface during rapid expansion of the cavity, in particular condensation.

The thermodynamic approach explains the role of the state of the interface in nucleation statistics without making any assumption about the microscopic structure of the cavity. An alternative bottom-up approach, for example assuming a physical cavity between the bilayer leaflets (37), has a major challenge in properly accounting for the energy required for the removal of interfacial water, which is closely tied to compression and condensation of the lipids to $L_c$ phase (31, 38–40). Our approach only asserts that nucleation is most likely at the site of maximum fluctuations, however it is not clear that the hydrophobic core would be that site and not the water-lipid interfacial region for example (41).

Similarly, during macroscopic expansion of the cavity, bottom-up models have intuitively assumed increase in surface tension upon expansion, however as discussed, cavitation induced dehydration would result in compression or increase in lateral pressure and hence a decrease in surface tension(26). Finally, apart from compression, our study also highlights the importance of the dehydration and jamming of lipids at the expanding interface. Such effects have also been shown to cause reduction in intra-lamellar distances and compression in MLV suspensions (35, 42), which has also informed our vision of the phenomenon.

Our vision of the cavitation phenomenon at the interface deduced from the observations reported here is sketched in Figure 8. The cavitation is preferentially nucleated at a lipid interface, due to increased fluctuations. Once formed, the cavity grows under the negative acoustic pressure and more lipid is incorporated into the interface. This lipid interface provides the enthalpy for vaporization of water molecules in to the cavity resulting in the lipid at the interface condensing to the $L_c$ gel phase.

We end by suggesting that there is no reason to assume that lipids are necessary for a condensation event to happen during the rapid expansion of a cavity. Similar thermodynamic considerations should also be relevant for the water-vapor interface during cavitation in the absence of lipids. In this regard, the present study may provide insight into the phenomenon of nucleation of ice during cavitation, also known as sono-crystallization, which has been proposed to explain experimentally observed nucleation of ice at the interface (43–45). Based on seminal work by Hickling (46) it is generally accepted that sono-crystallization is due to adiabatic compression by shock waves emitted from a bubble collapse and not from bubble expansion. At the time of Hickling's work there was only limited data on shock compression of water (47), and subsequent work has questioned the validity of a phase transition during shock compression (48) of up to several giga-pascals. We hypothesize that if the interface of a cavity in water is adiabatically uncoupled from the bulk then the interface will cool down as it needs to provide the enthalpy for vapor to form. A sufficiently large decrease in the enthalpy could result in the water forming ice at the interface.

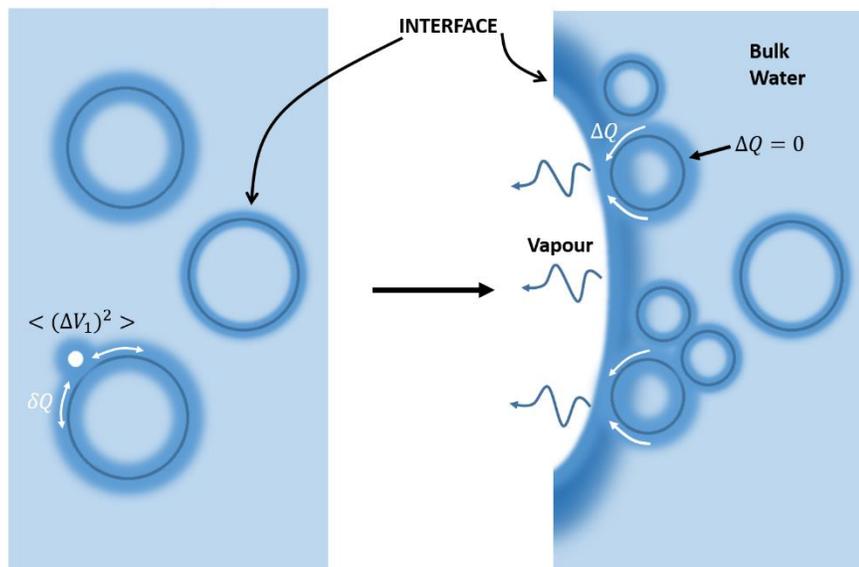

**Figure 8** Illustration of the proposed evolution of the interface during cavitation in a lipid environment. The interface is represented by the dense blue strip, the lighter shade the surrounding water, and white is the vapor cavity. (A) Due to reversible heat exchanges between lipids and interface water ($\delta Q$), microscopic vapor cavities are formed at the interface of a liposome (B) The initial cavity grows recruiting more lipids vesicles at the cavity interface; water will be lost from the lipid as vapor to the cavity, which requires heat of vaporization, which is compensated by the irreversible heat flow ($\Delta Q$) within the interface from lipids to water vapor (white arrows), whereas no heat flow from bulk to the interface (black arrow) due dynamic decoupling of the interface (see text), resulting in condensation of the interface.

The production of ice can be estimated by breaking down the adiabatic expansion into an isentropic and an isothermal (dissipation) process as follows. If we assume that the vapor pressure inside the cavity drops isentropically to 100Pa upon cavitation, then the temperature drops to approximately $T = -20°C$, as seen on a temperature-entropy (TS) diagram of water-vapor-ice (49). The entropy of vaporization of water is of the order of 6kJ/kg. K, so assuming a 15% rate of entropy production (represented by an isothermal process) the entropy change will be $s - s_0 \approx 0.9$kJ/kg. K, where $s_0$ is the entropy of water at 0°C and $100 kPa$. As can be seen on a water-vapor-ice TS state diagram (49), this amounts to approximately 4:1 ice to vapor mixture. Thus for every 5 water molecules entering the expanding boundary, 4 molecules will be deposited as ice at the interface (50) and 1 molecule will contribute to the volume of expanding cavity. Similar to relaxation of condensed lipids observed in this study, the ice should also disappear in millisecond time scales (51), as the pressure equilibrates to initial conditions and water returns to the equilibrium liquid state. However, if the initial state is that of supercooled water (metastable liquid), upon nucleation of microscopic ice during the negative pressure tail, the crystal will expand into the surrounding supercooled water as the system will relax to the new equilibrium state of a macroscopic mixture of ice and water, as has been observed during sonocrystalization experiments(52).

In conclusion, the results here suggest that the thermodynamics of interfaces plays a crucial role in the nucleation and subsequent evolution of cavitation. The cavitation threshold was found to be related to the magnitude of microscopic fluctuations in the interface, which is a function of the thermodynamic susceptibility of the interface - governed by the presence of lipids in this study. The threshold for cavitation was found to be minimum when the interface is close to phase transition as the fluctuations are most intense at that point. Once nucleated, if the cavity underwent a fast expansion, the interface was adiabatically uncoupled from the bulk and the lipids cooled and condensed to a gel phase as water formed vapor inside the cavity. In the case of a slower expansion the interface received sufficient heat from the bulk to become more fluid and remain hydrated. The critical rate which separates these two phenomenon was not resolved due to the limitations of current experimental setup. The study also has important implications for application emphasizing ultrasound and membranes, for example drug delivery, acoustic neuro-stimulation, and ultrasound contrast agents where the role interfacial entropy has not been incorporated in most models.

**Appendix A On fluctuations coupling at the interface**

In the following, we derive a relation explaining the observed minimum in the cavitation threshold (nucleation of vapor cavity in water) near a phase transition of the suspended lipids. Note that the derivation assumes nucleation results from reversible fluctuations, which is valid until the subsequent expansion of the cavity is irreversible. In homogenous nucleation theory, cavitation is regarded as water to vapor phase transition under tensile stress (53). Observed in the pressure-volume state diagram, the vapor phase becomes more stable once the pressure drops below the saturation boundary. However, the transition requires a finite time in order for the thermal fluctuations to overcome the entropy required to form the new interface (activation entropy, $\Delta S_{act}$). Therefore, if the system is subjected to tensile stress *quasi-statically, i.e.* the system has sufficient time to relax, the vapor phase appears at the saturation boundary. However, dynamic tensile stress, such as during a shock wave, can result in metastable states (region between the saturation line and the spinodal)(54). This is further complicated in the presence of lipids. As discussed below, a top-down approach is better suited to address such complexity as we have access to several macroscopic variables, starting from the shock wave induced pressure changes.

There are two main steps to understanding the effect of lipids on cavitation threshold in the presence of shock waves, (i) how are fluctuations in specific observables associated with water are affected by fluctuations in specific observables associated with the lipids at the interface (fluctuation coupling) (ii) how these fluctuations change in the presence of shockwaves.

**A.1 Fluctuations in stationary water**

Let us first consider stationary water without any lipids. From statistical thermodynamics the probability $w$ of a given state scales exponentially with the magnitude of entropy variation, i.e. $w = e^{\frac{(S-S_0)}{k}}$, where $S_0$ is the equilibrium entropy, $k$ is Boltzmann constant. The fluctuation in entropy is $|(\delta S)^2| \sim k C_p$, where $C_p$ is isobaric heat capacity (55). The meaning of fluctuation in entropy is not trivial, here it means variations in the entropy as an analytical function $S(x_1, x_2, ...)$ due to fluctuations in the variables $x_i's$, which, in principle, are experimentally observable, such as volume, concentrations, energy of a part of the system, and charge(56). Then $C_p$ measures the amount of heat that goes into changing $x_i's$ reversibly for a unit change in temperature at constant pressure. Similarly, volume fluctuations are $|(\delta V)^2| \approx k\beta_T VT$ where $\beta_T$ is isothermal compressibility. Cavitation becomes increasingly likely if the fluctuations are greater than the activation value, that is, $\sqrt{kC_p} > \delta S_{act}$ or equivalently $\sqrt{k\beta_T VT} > \delta V_{act}$. Figure 9 provides a visual representation of how these conditions affect nucleation statistics. While $|(\delta S)^2|$ represent variations on vertical axis, $|(\delta V)^2|$ result from the projection of the entropy potential $S(x_1, x_2, ...)$ on a horizontal axis, e.g. $x_1$, and $\beta_T$ is the projected curvature. Due to the Boltzmann principle, the system is exponentially more likely to be near the principle curvature, i.e. the system will on average fluctuate along the principle curvature. This is the reason why, near an equilibrium, it is not required to measure all other observables, i.e. $x_2, x_3, ...$, as $|(\delta S)^2|$ *on average* can be estimated based on any one of the $|(\delta x_i)^2|$ alone.

Note that $\delta V_{act}$ can be related to the critical radius $r_c$ in classical nucleation theory through $\delta V_{act} = \frac{4}{3}\pi r_c^3$, where $r_c = \frac{2\sigma(T)}{(P_l - P_v)}$ ($P_v$ is the vapor pressure and $P_l$ is the pressure in liquid) is of the order of nanometers and is typically not accessible as an experimental observable. In comparison, $\sqrt{k\beta_T VT}$ or $\sqrt{kC_p}$ are macroscopic properties of the entire system that can, in principle, be measured during an experiment. This provides a top-down approach to nucleation statistics, with a foundation ensured by the second law of thermodynamics (57). Note that in the presence of lipids, $\beta_T$ of the system undergoes a change and has a minimum near phase transition (58, 59), which is also the effective $\beta_T$ that should enter the fluctuation relation. Thus volume fluctuations in general will be maximized in the entire system, including those from (a) vapor cavities forming in water alone, (b) vapor cavities forming with lipids at the surface, (c) pressure volume changes in lipids or water in the absence of cavities. Thus even on the basis of $\beta_T$ it is implicitly clear that cavitation threshold will be lowered near phase transitions in the lipids. This can also be shown explicitly based on the coupling between the observables specific to lipids and water.

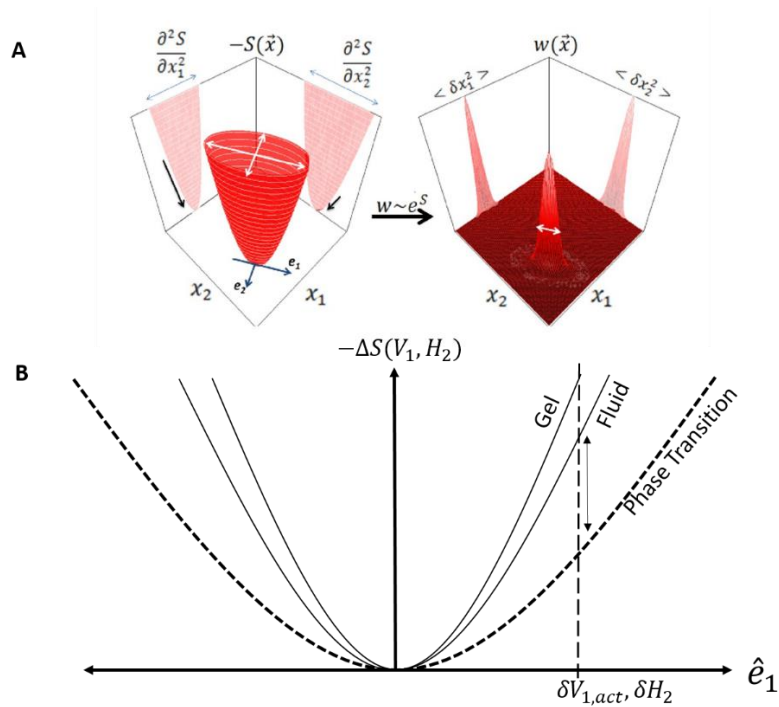

Figure 9. (A) Generalized quadratic entropy potential for two experimental observables $S(x_1, x_2)$ (60). The bases $x_1, x_2$ are not necessarily orthogonal however, corresponding $e_1, e_2$ can be found that are orthogonal along the principle curvatures of the quadratic potential. The projections indicate how the susceptibility of a particular observable is related to principle curvatures of the potential. The probability distribution is obtained from the entropy potential $w \sim e^S$, which gives fluctuations as the width of the distribution. Notice that for an arbitrary orientation of $e_1, e_2$ fluctuations as projected on $x_1, x_2$ will generally be coupled. (B) Projection of the potential on principal axis

$ê_1$, shows the effect of change in state in terms of change in the curvature of the potential (thermodynamic susceptibility). Near phase transition, the decreased curvature implies higher probability such that $\sqrt{|(\delta V)^2|} > \delta V_{act}$ and hence lowered threshold.

## A2. Fluctuations in stationary lipid and water system

When the system consists only of water, the macroscopic extensive variables have a simple meaning, which is directly related to the specific properties of water. In a mixture however this is not true as the macroscopic extensive variables will have contributions from different constituents. A typical approach is to represent them as a linear combination of the specific properties of the constituent, for example $S(x_i) = S_1(x_{i_1}) + S_2(x_{i_2}) = \alpha_1 s_1 + \alpha_2 s_2$ where $\alpha_i's$ can be the mole fractions. However this intrinsically assumes that the fluctuations in the two systems are independent, i.e. $w(x_1, x_2) = w(x_1).w(x_2)$, in other words it ignores the contribution of the interface $S_{1,2}$. Therefore, it is more rigorous to not assume any rule for the combination, rather a general functional dependence $S(x_1, x_2)$. Examples of entropy potential defined as a function of the parts of the system are rare (56), therefore, for the completeness and clarity of the derivation below it should be noted, for example ;

$$V = V_1 + V_2 \Rightarrow (\partial V)_{V_2} = (\partial V_1)_{V_2} \Rightarrow \frac{\partial S}{\partial V_1} = \frac{\partial S}{\partial V} = -\frac{P}{T}. \tag{A1}$$

$P$ is equilibrium pressure in the entire system and $T$ is the temperature. The physical significance of the partial derivatives with respect to the parts of the system can be inferred from a thought experiment where the gas is trapped between a piston and an incompressible fluid. Compressing the piston by $\partial V$ changes the entropy $\partial S$ of the system setting up a pressure through the entire system $\frac{\partial S}{\partial V} = -\frac{P}{T}$. However, the change in volume can primarily be traced to a change in the volume of the gas $\partial V_1$, which sets up the pressure $\frac{\partial S}{\partial V_1}$ in equilibrium with the rest of the fluid. Therefore, $\frac{\partial S}{\partial V_1} = \frac{\partial S}{\partial V} = -\frac{P}{T}$. This can be extended to other extensive observables as well, as long as the system is in equilibrium.

In our system, we have water, lipid membranes and dye molecules. The dye molecules act as a reporter of the interface and so can be considered as a single "membrane" system (7). Here we want to understand how the fluctuations in the observables specific to water (subsystem 1) are affected by the fluctuations in the observables specific to lipid membrane (subsystem 2). As seen in fig.9 and discussed above, the choice of $x_1, x_2$ (e.g. $V_1, V_2$) is arbitrary from a theoretical perspective. However, experimentally it should be based on what variables are accessible, given that choosing any $x_i$ does not affect the accounting of $S$. In our experimental setup $x_1$ is the volume of water (i.e. $V_1$ of subsystem 1) and $x_1$ is the enthalpy of lipids (i.e. $H_2$ of subsystem 2) are the corresponding representative variables. In which case expanding entropy potential of the entire system near an equilibrium yields;

$$S(V_1, H_2) = S_0 + \frac{1}{2}\frac{\partial^2 S}{\partial V_1^2}(\delta V_1)^2 + \frac{\partial^2 S}{\partial V_1 \partial H_2}\delta V_1 \delta H_2 + \frac{1}{2}\frac{\partial^2 S}{\partial H_2^2}(\delta H_2)^2 \tag{A2}$$

Making use of $\frac{\partial S}{\partial V_1} = \frac{\partial S}{\partial V}$ as shown in (A1) the mixed derivative can be expressed as,

$$\frac{\partial^2 S}{\partial V_1 \partial H_2} = \frac{\partial}{\partial H_2}\frac{\partial S}{\partial V_1} = \frac{\partial}{\partial H_2}\left(-\frac{P}{T}\right) = \frac{\frac{P}{C_{p_2}} - T\frac{\partial P}{\partial H_2}}{T^2} \tag{A3}$$

where $C_{p_2} = \frac{\partial H_2}{\partial T}$ is the heat capacity of the lipid. Comparing the coefficient with a Gaussian distribution of two variables corresponding to $w = e^{\frac{(S-S_0)}{k}}$, where $S$ is given by eq. A2, we get

$$|\delta V_1 \delta H_2| \approx \frac{kT^2}{P/C_{p_2} - T\frac{\partial P}{\partial H_2}} \tag{A4}$$

The relation shows how the enthalpy fluctuations $|(\delta H_2)^2|$ affect the volume fluctuations $|(\delta V_1)^2|$, which ultimately results in nucleation of vapor cavity. Furthermore, $|(\delta H_2)^2| \approx kC_{p_2}T^2$ which implies a maximum in $|(\delta H_2)^2|$ at the phase transition due to corresponding maximum in $C_{p_2}$. Recognizing that $\frac{\partial P}{\partial H_2}$ was previously measured to be negative and minimum during phase transition (enthalpy of lipids increased during negative phase of small pressure perturbation (7)), implies that the right hand side of eq. A4 is positive and maximum during phase transition. Hence $|(\delta V_1)^2|$ is maximum near a phase transition in lipids. The relation is valid as long as we can define heat capacity or compressibility locally as a function of dynamic pressure and temperature (i.e. the system is in local equilibrium), which is also going to allow their limited extension to shock wave induced state changes.

**A3 Fluctuations during a Shock Wave**

While the above discussion assumed complete equilibrium, it can be extended to an arbitrary macroscopic state (partial equilibrium) (61); by observing the system for a time interval $\Delta t$ that is small compared to the relaxation time for the equilibrium. The entropy is then the sum of entropies of subsystems so small that the corresponding individual relaxation times are much smaller than $\Delta t$. Then during $\Delta t$ each subsystem can be assumed in their particular local equilibrium. For example, in our experiment a time resolution of $1\mu s$ and speed of sound ~1500m/s gives the size of subsystem as ~1.5mm, which is of the same order as the FOV of the optical setup. Furthermore, $1\mu s$ is significantly smaller than the cavitation timescale as well as lateral diffusion timescales of the lipids. Indeed, we have previously shown that for subthreshold acoustic impulses of up to EL 5, the assumption of local equilibrium correctly predicts the response of lipids. That is, the acoustic response of lipids for subthreshold pressure pulses was maximum at phase transition where the heat capacity is maximum (7).

Thus, small acoustic perturbations can be imagined as the system oscillating in the entropy potential of fig.9 (62). However, increasing the amplitude can then deform the potential itself (curvatures, i.e. heat capacity and compressibility, vary as a function of pressure). Indeed, as the shock wave expands the liquid into the metastable regime, compressibility increases and diverges as the liquid approaches vaporization spinodal (63–65), and since $|(\delta V)^2| \approx k\beta_T VT$, fluctuations diverge as well making nucleation unavoidable. The presence of lipids further contributes to these fluctuations as discussed thus lowering the threshold for nucleation. Beyond nucleation the system becomes unstable resulting in irreversible *macroscopic* expansion of a cavity, due to non-equilibrium heat transfer and the equations above are not valid anymore.

**Acknowledgement**

We acknowledge the support by Engineering and Physical Sciences Research Council (EPSRC) under Programme Grant EP/L024012/1 (OxCD3: Oxford Centre for Drug Delivery Devices). We thank Prof. Lawrence Crum for a critical reading of the manuscript. Data will be made available on Oxford Research Data Archive http://researchdata.ox.ac.uk/

# Supplementary Information

# Thermodynamic State of the Interface during Cavitation

**Shamit Shrivastava and Robin O. Cleveland**

Department of Engineering Science, University of Oxford, UK

**Supplementary Figures**

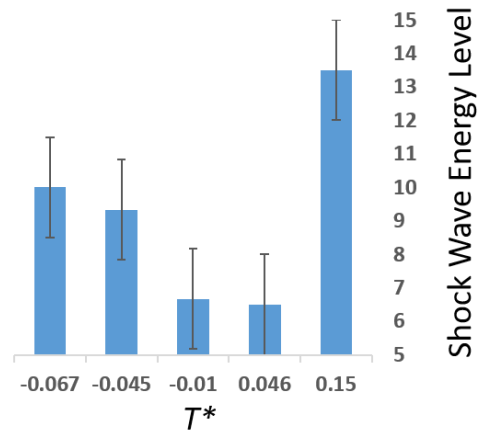

**SI Figure 1** The mean minimum energy level at which >50% cavitation was observed in the optical signal as a function of $T^*$. The minimum energy level setting are required when $T^* \sim 0$, that is when the lipids are at the transition temperature. The overlaying bars of magnitude 1.5 energy levels indicate the magnitude of discrete gap between energy levels of the shockwave source.

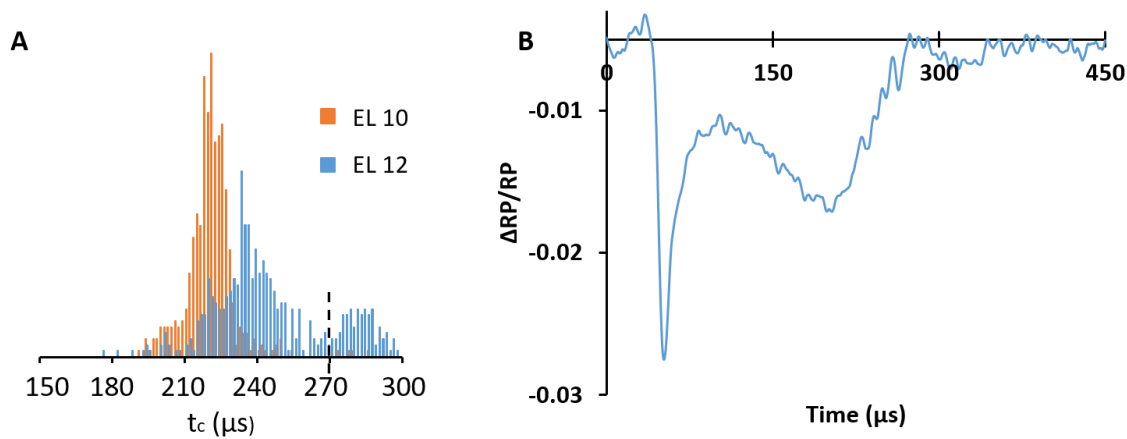

**SI Figure 2** (A) The histogram for cavitation times at setting 10 and 12. The bimodal distribution at EL 12 was used to separate waveforms with $t_c > 270 \; or \; < 270 \mu s$ to produce the two waveforms in fig 6A and 6B respectively. (B) shows the average waveform for entire populations to showing that, if observed, such average waveforms need to be properly decomposed.

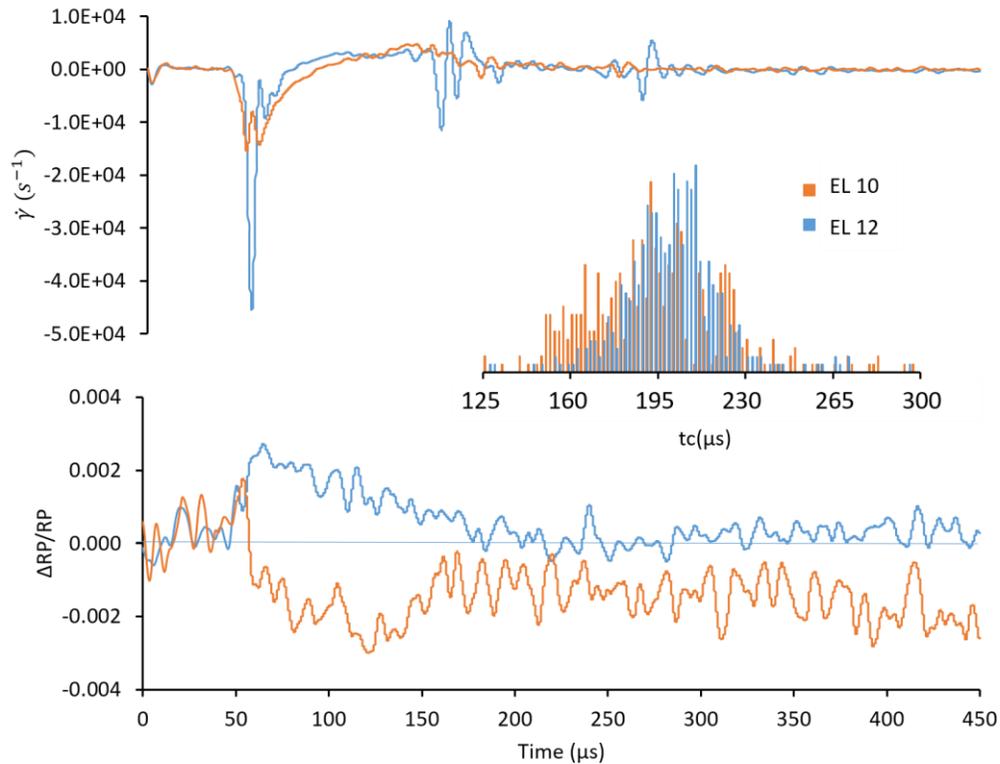

**SI Figure 3** Expansion rate dependence of cavitation induced state changes in DMPC suspension at 10°C, i.e. $L_\beta$ gel phase. (Top) Expansion rate $(-\dot{\gamma})$ in two different preparations of DMPC suspension at setting 10 and 12. Inset shows distribution of collapse time in the two experiments is very similar. Yet, ΔRP/RP change (Bottom) corresponding to two different expansion rates in two different experiments is qualitatively different. Higher expansions rate results in condensation while lower rate in fluidization.

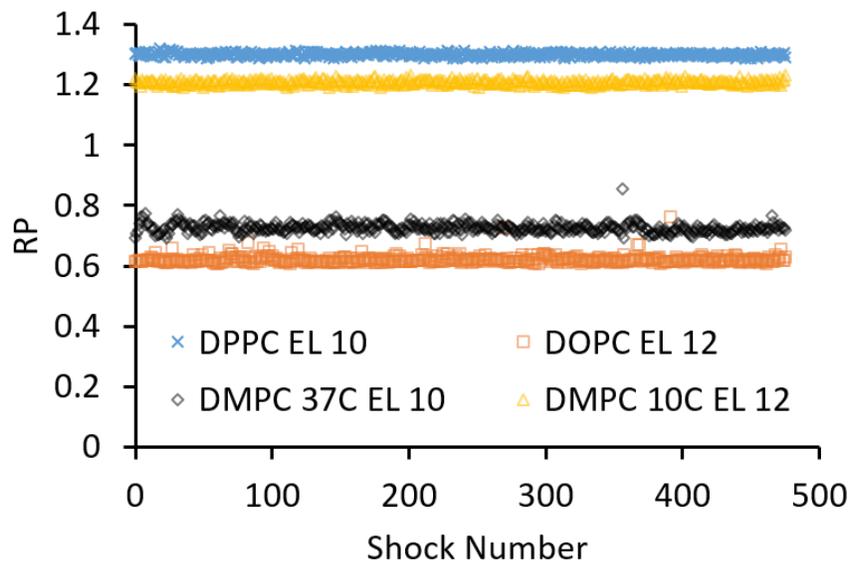

**SI Figure 4** The mean RP measured before every shock during the course of an experiment shows that the cavitation induced state changes are completely reversible, with some variability in DPMC at 37C, likely due to increased sensitivity of RP close to the transitions.